\documentclass[11pt]{article}
\usepackage{moriond,epsfig}

\def\Journal#1#2#3#4{{#1} {\bf #2}, #3 (#4)}

\def\be{\begin{equation}}
\def\ee{\end{equation}}
\def\bea{\begin{eqnarray}}
\def\eea{\end{eqnarray}}

\def \alf{\alpha_{\rm ph}}

\begin{document}
\vspace*{4cm}
\title{THE SPECTRUM OF THE MILLISECOND PULSAR J0218$+$4232 -- 
THEORETICAL INTERPRETATIONS}

\author{ J. DYKS, B. RUDAK }

\address{Nicolaus Copernicus Astronomical Center, Rabia\'nska 8,\\
87--100 Toru\'n, Poland}

\maketitle\abstracts{
  We interpret the unique high-energy spectrum 
  of the millisecond pulsar PSR J0218$+$4232
  within polar cap scenarios.
  We show that 
  the spectral data from BeppoSAX \cite{mck2000}  and 
  \hbox{EGRET \cite{khv2000}}
  impose very restrictive limitations on possible radiation mechanisms,
  energy spectrum of radiating charges 
  as well as viewing geometry.
  Theoretical spectra are able to reproduce
  the data, however,
  this can be achieved provided very special 
  -- unusual within the conventional polar cap picture -- conditions are
  satisfied.
  Those include off-beam viewing geometry along with
  one of the following alternatives:
  1) strong acceleration of secondary pairs;
  2) broad energy distribution of primary electrons extending down 
     to $10^5$ MeV;
  3) high-altitude synchrotron emission.
}

\section{Introduction}

PSR J0218$+$4232,
with the spin period $P = 2.3$ ms and the inferred dipolar magnetic field at
polar cap $B_{\rm pc} \simeq 8.6\times 10^8$ G, 
is the only millisecond pulsar which has been
marginally detected above 100 MeV \cite{vkb1996}.
A broad-band high-energy spectrum of this object exhibits unusual
features, completely different from those observed among young gamma-ray 
pulsars:
Above 100 MeV the photon index $\alf \sim -2.6$ \cite{khv2000}
and the spectrum 
resembles the very soft spectra of middle latitude
unidentified EGRET sources.\cite{gp2001}
Within the BeppoSAX range the spectrum is extremely hard: 
$\alf \simeq -0.61\pm 0.32$.\cite{mck2000}

Kuiper et al.\cite{khv2000} noticed that neither 
the polar cap nor the outer gap
models could naturally explain the spectrum. 
The aim of this work is to 
carefully examine an ability of polar cap models to reproduce the spectrum
of J0218$+$4232.

\section{Directional characteristics of pulsar spectra
predicted by the polar cap model}\label{directional}

As we show in the acompanying paper (Wo\'zna et al., these proceedings)
the viewing geometry effects have crucial significance for the appearance
of a pulsar spectrum for a given observer. 
For purely dipolar magnetic field the geometry is determined by two
angles: $\alpha$ -- the angle of dipole inclination relative to the rotation
axis, and $\zeta$ -- the angle between an observer's line of sight and
the rotation axis. 
Various combinations of $\alpha$ and $\zeta$ result in a large variety
of spectral shapes and pulse profiles in high-energy domain. 
However, a particular energetic history of electrons
in pulsar magnetosphere enables to extract two main cases: the on-beam and
the off-beam geometry.

\begin{figure}[t]
\begin{minipage}[l]{.40\textwidth}
 \includegraphics[width=\textwidth]{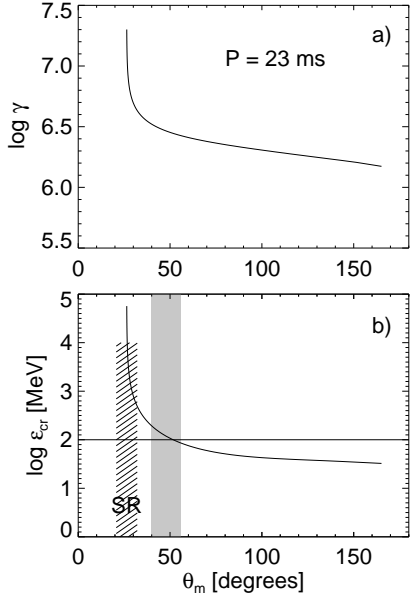}
\end{minipage}%
\hfill
\begin{minipage}[r]{.60\textwidth}
\vskip-6mm
\caption{ a) Electron energy $\gamma$ as a function 
of magnetic colatitude of
photon propagation direction $\theta_{\rm m}$ for $P=23$ ms and $\gamma_0 =
2\times 10^7$; b) Characteristic energy of curvature photons $\epsilon_{\rm
cr}$ as a function of $\theta_{\rm m}$ for the electron energy shown in the 
upper panel. 
The dashed region (SR) indicates the range of $\theta_{\rm m}$ where most of
synchrotron photons are emitted.
The figure implies that for most viewing angles $\zeta$ one of two 
main shapes of spectrum can be recorded: 
In the case of the on-beam geometry
the line of sight samples 
$\theta_{\rm m} < 1.5\ \theta_{\rm pc}+1/\gamma_\parallel \simeq 32^\circ$
and the observed spectrum consists of curvature and synchrotron components.
The first one extends up to $\sim 10^5$ MeV with  
$\alf \sim -1.66$ and the latter one, with $\alf = -1.5$,
dominates below a few tens of MeV.
For the off-beam geometry only the curvature component can be detected 
with $\alf = -2/3$ below $\sim 10$ MeV and $\alf >
2$ within the range of its high-energy cutoff around $100$ MeV. 
PSR B1821$-$24 and PSR J0218$+$4232 seem to be the examples
of these two cases, respectively. An intermediate case - with the line of
sight just grazing the heighest-energy gamma-ray beam - is less probable. 
The gray band marks the range of
$\theta_{\rm m}$ sampled by the line of sight at $\zeta = 48^\circ$ for
$\alpha = 8^\circ$. The spectrum corresponding to this off-beam viewing geometry 
is shown as a thick solid
line in Fig.~2. 
}
 \label{fig1}
\end{minipage}
\end{figure}

Fig.~1a presents the Lorentz factor $\gamma$ of electrons injected at the
surface of neutron star along the "last open" magnetic field lines as a
function of angle $\theta_{\rm m}$ between magnetic dipole axis and a
local tangent to magnetic field line at the electrons' position.
The value of $\theta_{\rm m} = 26.^\circ 4 \simeq 1.5\ \theta_{\rm pc}$ corresponds 
to the direction of $\vec B$ at the
polar cap rim ($\theta_{\rm pc}$ is the angular radius of the polar cap,
measured from the center of neutron star). 
The rotation period $P= 2.3$ ms and the initial Lorentz factor 
$\gamma_0 = 2\times
10^7$ were assumed in these calculations.
The electron energy losses noticeable in Fig.~1a are purely due to the emission
of curvature radiation \footnote{Resonant inverse Compton scatterings cannot 
influence the
electron's energy due to
small relative energy losses per scattering in a weak magnetic field
(eg.~Dyks, Rudak, Bulik~\cite{drb2001}) 
whereas non-resonant
scatterings (they occur in the Klein-Nishina regime) are too rare for
surface
temperature values relevant for neutron stars.\cite{brd2000}
}
The CR energy loss rate of an electron is initially huge, but 
because of its strong dependence on electron
energy ($\dot\gamma_{\rm cr} \propto \gamma^4/\rho_{\rm cr}^2$,
$\rho_{\rm cr}$ is the local radius of curvature of magnetic field lines)
it becomes negligible
after the electron traverses a length which is small in comparison with the
light cylinder radius $R_{\rm lc}$. Thus, after the initial rapid drop,
the electron's energy starts to decrease very slowly (Fig.~1a).

The characteristic energy $\epsilon_{\rm cr}$ of curvature photons emitted 
by the electron 
in different directions $\theta_{\rm m}$ is shown in Fig.~1b.
Because $\epsilon_{\rm cr}$ is very sensitive to $\gamma$ 
($\epsilon_{\rm cr}\propto
\gamma^3/\rho_{\rm cr}$)
the high-energy cutoff in the CR spectrum
decreases rapidly for increasing angles $\theta_{\rm m}$ but then it starts
to approach "asymptotically" the value of $\sim
30$ MeV. This is in part due to the slower decrease in $\gamma$ at higher
altitudes and in part
due to an increase in $\rho_{\rm cr}$ which starts to take place 
for $\theta_{\rm m} > 68^\circ$. In fact, Rudak \& Dyks \cite{rd1999} (RD99)
estimated an "absolute minimum" of $\gamma_{\rm break} \simeq 3.8\times 
10^6 (\rho_{\rm cr}/(10^7\ {\rm
cm}))^{1/3}$ for the Lorentz factor of CR-cooled
electrons (eq.~(7) in RD99) and the absolute lower limit of $\sim 150$ MeV 
for $\epsilon_{\rm cr}$ (eq.~(8) in RD99).
These values are in reasonable agreement 
with the exact results presented in Fig.~1, given the crude method used in
RD99. We recall here that the lower
limit for $\epsilon_{\rm cr}$ practically does not depend on 
any pulsar parameters
(eg.~exact calculations for rotation period $P = 0.1$ s give $
\epsilon_{\rm cr} \sim 20$ MeV at $R_{\rm lc}$).

An important implication of Fig.~1b is that pulsar radiation pattern can be
considered as consisting of two components: a hollow cone of very high-energy
emission extending up to GeV range (with opening half-angle $\theta_{\rm m}
\simeq 1.5\ \theta_{\rm pc}$)
and a much less anisotropic emission peaking close to 30
MeV. 
When the line of sight 
crosses the hollow cone beam (on-beam geometry)
the recorded spectrum consists of two components: a curvature component and a
synchrotron component due to synchrotron radiation  (SR) from secondary
electron-positron pairs. The shape of this spectrum is similar to the total
spectrum emitted in all directions, shown in fig.~1 of RD99.
Its qualitative features can also be assesed from Fig.~1b: The spectrum
extends  
up to GeV range and is relatively soft within the EGRET range
because it is composed of instantaneous
CR spectra with different values of $\epsilon_{\rm cr}$
(Fig.~1b). The value of photon index $\alf$ in this spectral range 
depends on the viewing angle $\zeta$  as well as on the rotation period $P$. 
For millisecond periods it is close to $\alf \sim -5/3$.
Within the entire X-ray range ($0.1$ keV -- a few MeV) the on-beam spectrum 
is dominated by the synchrotron component
with the well-known photon index $\alf = -1.5$ (RD99).

As noted by Kuiper et al.\cite{khv2000}, these on-beam
 characteristics are in clear
disagreement with the high-energy data on J0218$+$4232.
We, therefore, propose the off-beam 
geometry for this pulsar ie.~the case
when the line of sight misses the hollow cone beam.
With this assumption
the problem of wrong slopes within the BeppoSAX and the EGRET
range can be solved.\\
Since CR photons of the highest-energy are converted into the $e^\pm$ pairs 
with Lorentz factors $\gamma_\parallel \sim 10$ (in $B\sim 10^9$ G), 
most of synchrotron photons
follow the direction of the "parent" CR photons and are constrained to a
narrow range of angles around $1.5\ \theta_{\rm pc}$ with a spread of
$1/\gamma_\parallel \simeq 0.1$ sr (Fig.~1b).
Therefore, 
in the off-beam case one misses the SR emission and the spectrum
consists of the CR component only. Thick solid line in Fig.~2 presents 
this kind of spectrum calculated for $\alpha = 8^\circ$ 
\cite{stc1999} 
and $\zeta = 48^\circ$ (which corresponds to $40^\circ < \theta_{\rm m}
< 56^\circ$ in Fig.~1).
The spectrum resembles very closely
the well-known instantaneous CR spectrum 
due to monoenergetic electrons with no cooling.
The reason for this is clearly shown in Fig.~1b:
because of the near alignment of the dipole and the rotation axis
($\alpha \simeq 8^\circ$) the off-beam line of sight samples only a very
limited range of $\theta_{\rm m}$: between $40^\circ$ and $56^\circ$
(grey band in Fig.~1b).
\footnote{
Interestingly, the shape of the spectrum would not change much even if
the line of sight sampled much larger range of $\theta_{\rm m}$ within 
the off-beam region (say between $40^\circ$ and $140^\circ$). This is because
of the remarkable stability of $\epsilon_{\rm cr}$ at high altitudes -- see
Fig.~1b.}
Below a high-energy cutoff at $\sim 100$ MeV
the CR spectrum has the photon index $\alf = -2/3$.

\begin{figure}[t]
\begin{minipage}[l]{.68\textwidth}
 \includegraphics[width=\textwidth]{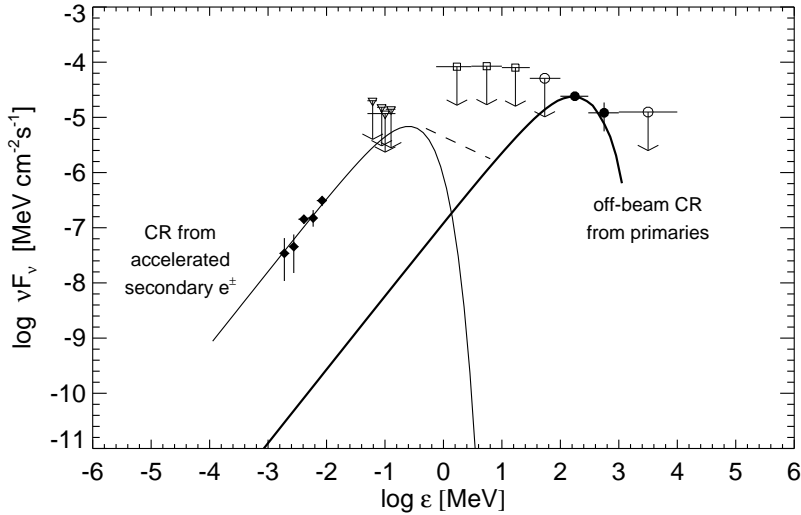}
\end{minipage}%
\hfill
\begin{minipage}[r]{.3\textwidth}
\vskip-6mm
\caption{Comparison of theoretical pulsar spectrum for the off-beam geometry
(thick solid line) with the spectrum observed for PSR J0218$+$4232
(diamonds -- BeppoSAX, triangles
-- OSSE, squares -- COMPTEL, circles -- EGRET). 
The BeppoSAX points suggest a presence of 
additional spectral 
component in the X-ray domain.
Such a component would arise due to CR of secondary $e^\pm$ pairs, 
provided that at least a few hundred of
them per primary are first 
accelerated up to $\gamma \sim 10^5$ MeV (see Section 3).  
}
 \label{fig2}
\end{minipage}
\end{figure}

If the X-ray emission from J0218$+$4232
is generated close to the neutron star surface, it can be
explained only as a pure CR emission (note that $\alf = -2/3$ matches well
the BeppoSAX photon index $-0.61\pm0.32$).
Any SR emitted close to the surface would have a broken power-law shape with
$\alf = -1.5$ within the X-ray range, in clear disagreement with the
BeppoSAX data.
The slope $\alf = -2/3$ of the CR component
does not depend on the viewing angle $\zeta$
as long as the line of sight misses the SR component, (which is 
concentrated around
$1.5\ \theta_{\rm pc}$),  ie.~as long as $\zeta > \alpha + 
1.5\ \theta_{\rm pc} + 1/\gamma_\parallel \simeq 40^\circ$
(see Wo\'zna et al., this volume).

Furthermore, owing to the off-beam geometry
the high energy cutoff in the total spectrum (which now consists 
of just a CR component) occurs at a relatively low photon energy 
of $\sim 100$ MeV.
The cutoff is not caused by magnetic absorption -- it corresponds to
a maximum energy of those electrons which emit observable CR. 
This is why the cutoff's shape can easily 
mimic the very soft spectral shape as suggested by the two EGRET points
(Fig.~2).
Harding \& Zhang \cite{hz2001} used similar viewing geometry arguments
to explain the very soft spectra of unidentified EGRET sources.

However, with the off-beam CR spectrum normalized to reproduce 
the level of the EGRET emission, 
the BeppoSAX data points are located about 3 orders of 
magnitude above the
extrapolated level of the CR component
(thick line in Fig.~2).
The intrinsically hard shape of the off-beam spectrum results in too low
ratio of X-ray level to gamma-ray level of emission. 
In our opinion it is not possible
to solve this problem within the standard polar cap model.

\section{Can the secondary $e^\pm$ pairs emit CR photons within the keV range?} 

In order to explain the BeppoSAX data we propose
a contribution of CR from secondary $e^\pm$ pairs accelerated 
within the polar gap. 
Upon the acceleration energy $\gamma_\parallel$ of 
the $e^\pm$ pairs should reach values in the range between $1.5\times10^5$ and
$6\times 10^5$. For $\gamma_\parallel < \gamma_{\rm break} \sim
3\times10^6$ the CR energy loss length scale considerably exceedes the size
of magnetosphere ($R_{\rm lc}$). The pairs do not suffer thus considerable 
energy losses and at all altitudes emit roughly the same spectrum of CR.
In Fig.~2 it is shown as a thin solid line (we took
$\gamma_\parallel = 3\times 10^5$).

The level of CR spectrum at energies below the characteristic energy
$\epsilon_{\rm cr}\propto \gamma_\parallel^3/\rho_{\rm cr}$
does not depend on $\gamma_\parallel$, but solely on the radius of curvature
of magnetic field lines $\rho_{\rm cr}$. 
Since the pairs follow nearly the
same field lines as the primary electrons, 
the ratio betwen 
the level of BeppoSAX data points (diamonds in Fig.~2) and 
the level of X-ray
emision from primary electrons (thick solid line)
is a direct measure of the number 
of pairs $n_\pm$ per single primary electron required 
to reproduce the level of the X-ray data.
The level of the primary CR spectrum (thick solid line) within the BeppoSAX
range can be determined by fitting the high-energy cutoff of the 
spectrum to the EGRET data points. Apart from $n_\pm$, 
the same fit could also determine the value of $\zeta$, because
the shape of the cutoff depends on the viewing angle.\footnote{
Because of the near alignment of the magnetic dipole, the value of
$\zeta$ cannot be well determined from radio polarisation data on 
J0218$+$4232 (Stairs et al.\cite{stc1999})},
however, the EGRET data do not allow to determine 
$n_\pm$ and $\zeta$ with high accuracy.
"By eye" fits give $n_\pm$ of at least a few hundred; 
for $\zeta = 48^\circ$ (Fig.~2), $n_\pm \sim 10^3$.

Thus, to reproduce the broad-band high-energy spectrum
of J0218$+$4232, at least a few hundred of pairs should  
acquire
an energy roughly equal to $1\%$ of primary electron energy.
According to most works on the physics of the
pair formation front (eg. Harding \& Muslimov \cite{hm2001})
acceleration of such a large number of pairs within the polar gap is
difficult, because redistribution of the pairs screens out
the electric field of the polar gap.
However, existing works on the physics of the polar gap 
neglect the fact that
the $e^\pm$ pairs are created at different magnetic field lines than 
those along which parent primary electrons propagate. Because of this 
1D-approach as well as plenty of other approximations
our present knowledge of processes at the pair formation front
is far from being well established. 

Because of the bimodal energy distribution of radiating charges
(primary electrons, secondary pairs), the CR spectral components which
correspond to them are separated by a dip.
A precise shape of this dip would depend on the shape of
high-energy cutoff in the energy distribution of $e^\pm$ pairs
and the dip may extend in principle between $\sim 10$ keV
and $100$ MeV (dashed line in Fig.~2 suggests a possible situation).

\section{What is the energy spread among primary electrons?}

One of popular assumptions present in polar cap models is that all 
primary electrons
accelerated within the gap acquire roughly the same energy.
This is not realistic because the potential drop across 
the polar cap depends on
magnetic colatitude: it decreases towards the rim of the polar cap
(eg. \cite{dr2000a}$^,$ \cite{hm2001}).
Moreover, stochastic energy losses caused by resonant Compton scatterings may
significantly soften the energy spectrum of primary electrons
provided the surface temperature $T$ and the magnetic field $B$
are sufficiently high (see fig.4 in paper \cite{dr2000b}). In the case of
weak-$B$ millisecond pulsars the latter mechanism is not efficient
because the energy loss rate due to the resonant ICS ($\propto B$) is low.
However, no calculations of primary electron energy distribution
have been performed so far for very high
temperatures in excess of $10^7$ K -- a value considered for 
J0218$+$4232.\cite{mck2000}

Suppose that the primary electrons assume 
the steady-state energy distribution
in a form of a power law 
$dN_{\rm e}/d\gamma \propto \gamma^p$
extending down to $\gamma_{\rm min} = {\rm a\ few}\times 10^5$.
Then, 
the resulting CR spectrum could easily reproduce both the observed 
slopes and levels of X-ray and gamma-ray emission for appropriate values of
$p$ and $\gamma_{\rm min}$.
It would have a broken power law
shape with $\alf = -2/3$ within the BeppoSAX range,
and $\alf = (p - 1)/3$ above a break somewhere between 10 keV and 100 MeV
(the exact value of the break energy $\epsilon_{\rm br} = \epsilon_{\rm
cr}(\gamma_{\rm min})$).
Again, 
if primary electrons really achieved $\gamma$ in excess of $10^7$
in the polar gap of millisecond pulsars,
the agreement of
the model spectrum with EGRET data points could be easily achieved only by
a proper choice of the viewing angle $\zeta$ for the off-beam geometry.
There would be no dip in the spectrum
and a measure of  $\alf$ and $\epsilon_{\rm br}$ would give 
us direct information about $p$ and $\gamma_{\rm min}$, respectively.
BeppoSAX data points and COMPTEL upper limits constrain $\alf$ to the
range between $-1.5$ and  $-2.15$ which corresponds to rather soft
electron energy distribution with the index $-3.5 > p > -5.5$ and with the low energy
cutoff at $10^5 < \gamma_{\rm min} < 2.5\times 10^5$.

\section{High-altitude synchrotron emission}

In the case when the synchrotron radiation (SR) is emitted close to the
neutron star surface, its spectrum extends down to a blue-shifted
local cyclotron energy $\gamma_\parallel \hbar \omega_B$ with $\alf =
-1.5$ (RD99). For high-altitude emission, however, the cooling length scale
due to SR can become longer than the length scale of decrease in $B$.
Simple calculations show that this may happen for a local magnetic field
$B <
2\times 10^6$ G.
For PSR J0218$+$4232 such a field is expected at $r > 0.5\ R_{\rm lc}$.

In the case of SR in such a low $B$, electrons do not loose their entire
energy $\gamma_\perp mc^2$ corresponding to their motion across $B$.
In consequence, a new break appears in the SR spectrum, at photon energy
greater than $\gamma_\parallel \hbar \omega_B$ (Chang et al.\cite{cgl1999}).
Below the break, the photon index assumes the value $\alf \simeq -2/3$,
characteristic for instantaneous spectrum of SR (or CR) emission.

Fig.~3 shows that such a kind of spectrum is also able to reproduce the
high-energy data on J0218$+$4232. Between the BeppoSAX and EGRET range the
SR spectrum
has a well known slope of $\alf = -1.5$,
and its flux level is about 10 times  below the upper limits from OSSE and
COMPTEL.
This SR spectrum was calculated within the following simplified model: 
The electrons were
injected at a distance of $0.8 R_{\rm lc}$ from a neutron star,
with initial energy $\gamma = 3\times 10^7$ 
and were propagated radially up
to $5 R_{\rm lc}$ with a constant value of $\gamma_\parallel = 10^5$
(ie.~constant pitch angle during the SR emission was asumed).
Viewing geometry effects were not taken into account in these simple
calculations but the lack of strong GeV emission from J0218$+$4232
again implies the off-beam case.
The agreement between the data 
and the model spectrum of SR shown in Fig.~3 is remarkable, 
however, the emission of 100 MeV
synchrotron photons close to the light cylinder requires very high energy of
primary electrons near $R_{\rm lc}$: ($\gamma_\perp > 10^2$,
$\gamma_\parallel \sim 10^5$).
Various mechanisms of acquiring $\gamma_\perp > 1$ in the outer parts of
magnetosphere
have been considered so far within the polar cap model 
(eg.~Malov \& Machabeli \cite{mm2001})
but they usually concerned
the optical emission from pulsars.

\begin{figure}[t]
\begin{minipage}[l]{.68\textwidth}
 \includegraphics[width=\textwidth]{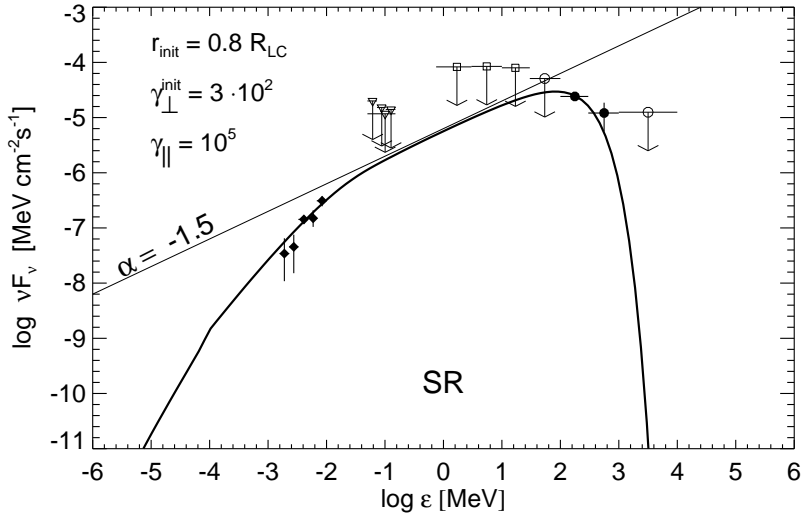}
\end{minipage}%
\hfill
\begin{minipage}[r]{.3\textwidth}
\vskip-6mm
\caption{Comparison of theoretical pulsar spectrum for a pure SR emission
close to (and beyond) the light cylinder 
(thick solid line) with the spectrum observed for PSR J0218$+$4232.
Note that the characteristic slope of SR emission
with $\alf = -1.5$ for $\epsilon > 30$ keV changes at $30$ keV 
into $\alf \simeq -2/3$ within the BeppoSAX range.  
}
\label{fig3}
\end{minipage}
\end{figure}

\section{Conclusions}

The combined X-ray and gamma-ray data on the millisecond pulsar J0218$+$4232
impose severe limitations on possible radiation mechanisms, energy
spectrum of radiating charges as well as viewing geometry.
The {\em standard} polar cap model is not able to explain the spectrum of     
J0218$+$4232
even when viewing geometry effects are taken into account.
We find three possible interpretations which require non-orthodox
assumptions about the electron energy distribution or emission altitude.
The spectra corresponding to these possiblities have unique features
(the MeV dip or the characteristic slope of SR) 
which may enable to identify them with 
a high-sensitivity and high-angular resolution gamma-ray telescope
(INTEGRAL?).

\section*{Acknowledgments}
We thank W. Hermsen and L. Kuiper for numerous discussions on J0218$+$4232.
BR acknowledges warm hospitality in SRON (Utrecht)
where this project was started.
JD acknowledges Young Resercher Scholarship of
Foundation for Polish Science. 
This work was supported by KBN grants 2P03D 02117 and 5P03D 02420.


\section*{References}
\begin{thebibliography}{99}
\bibitem{brd2000}T. Bulik {\it et al}, \Journal{MNRAS}{317}{97}{2000}
\bibitem{cgl1999}H.-K. Chang {\it et al}, \Journal{Astrophysical Letters \&
Communications}{38}{53}{1999} 
\bibitem{dr2000a}J. Dyks and B. Rudak, \Journal{A\&A}{362}{1004}{2000}
\bibitem{dr2000b}J. Dyks and B. Rudak, \Journal{A\&A}{360}{263}{2000}
\bibitem{drb2001}J. Dyks {\it et al.} \Journal{in Proc. of the 4th INTEGRAL
Worshop, ESA SP}{459}{191}{2001}
\bibitem{gp2001}I. Grenier and C. Perrot,
\Journal{AIP Conference Proceedings}{587}{649}{2001} 
\bibitem{hz2001}A.K. Harding and B. Zhang, \Journal{ApJ}{548}{L37}{2001} 
\bibitem{hm2001}A.K. Harding and A.G. Muslimov, \Journal{ApJ}{556}{987}{2001} 
\bibitem{khv2000}L. Kuiper {\it et al}, \Journal{A\&A}{359}{615}{2000}
\bibitem{mm2001}I.F. Malov and G.Z. Machabeli, \Journal{ApJ}{554}{587}{2001} 
\bibitem{mck2000}T. Mineo {\it et al}, \Journal{A\&A}{355}{1053}{2000} 
\bibitem{rd1999}B. Rudak and J. Dyks, \Journal{MNRAS}{303}{477}{2000} (RD99)
\bibitem{stc1999}I.H. Stairs {\it et al}, \Journal{ApJSS}{123}{627}{1999} 
\bibitem{vkb1996}F. Verbunt {\it et al}, \Journal{A\&A}{311}{L9}{1996} 
\end{thebibliography}
\end{document}